# Reversible Al Propagation in Si$_x$Ge$_{1-x}$ Nanowires

Luong Minh Anh[1], Robin Eric[1], Pauc Nicolas[2], Gentile Pascal[2], Baron Thierry[3], Salem Bassem[3], Sistani Masiar[4], Lugstein Alois[4], Spies Maria[5], Fernandez Bruno[5], den Hertog Martien[5]*

While reversibility is a fundamental concept in thermodynamics, most reactions are not readily reversible, especially in solid state physics. For example, thermal diffusion is a widely known concept, used among others to inject dopant atoms into the substitutional positions in the matrix and improve the device properties. Typically, such a diffusion process will create a concentration gradient extending over increasingly large regions, without possibility to reverse this effect. On the other hand, while the bottom up growth of semiconducting nanowires is interesting, it can still be difficult to fabricate axial heterostructures with high control.  In this paper, we report a reversible thermal diffusion process occurring in the solid-state exchange reaction between an Al metal pad and a Si$_x$Ge$_{1-x}$ alloy nanowire observed by in-situ transmission electron microscopy. The thermally assisted reaction results in the creation of a Si-rich region sandwiched between the reacted Al and unreacted Si$_x$Ge$_{1-x}$ part, forming an axial Al/Si/Si$_x$Ge$_{1-x}$ heterostructure. Upon heating or (slow) cooling, the Al metal can repeatably move in and out of the Si$_x$Ge$_{1-x}$ alloy nanowire while maintaining the rod-like geometry and crystallinity, allowing to fabricate and contact nanowire heterostructures in a reversible way in a single process step, compatible with current Si based technology. This interesting system is promising for various applications, such as phase change memories in an all crystalline system with integrated contacts, as well as Si/Si$_x$Ge$_{1-x}$/Si heterostructures for near-infrared sensing applications.

---

[1]Université Grenoble Alpes, CEA-Grenoble, IRIG-DEPHY-MEM-LEMMA, F-38054 Grenoble, France. [2]Université Grenoble Alpes, CEA-Grenoble, IRIG-DEPHY-PHELIQS-SINAPS, F-38000 Grenoble, France. [3]Université Grenoble Alpes, CNRS, LTM, 38054 Grenoble, France. [4]Technische Universität Wien, Institute of Solid State Electronics, Gußhausstraße 25-25a, Vienna 1040, Austria.  [5]Université Grenoble Alpes, CNRS, Institut NEEL UPR2940, 25 Avenue des Martyrs, Grenoble 38042, France.

*«martien.den-hertog@neel.cnrs.fr»

## Introduction

Group-IV semiconducting nanowires (NWs) are widely studied both by top down as well as bottom up approaches[1,2]. However, fabricating and contacting complex heterostructures with well-defined interfaces and contacts is still challenging, for example due to the reservoir effect of the catalyst particle for bottom up grown NWs[3–6]. Moreover, contacting of NWs is likewise still challenging. Several papers have shown that a metal propagation in the NW can be used to form an intermetallic phase with a very abrupt contact with the original NW[7–10]. Having the metal contact inside the NW geometry has the additional advantage that the contacts do not screen the remaining semiconducting region from the gate[11], as is the case with large metal contacts.

The diffusion behavior of Si/Al and Ge/Al binary systems in bulk and thin film materials has been studied for a long period resulting in a large amount of publications. In the Si/Al couple, due to the low solubility limit of Si in Al (about 1.62%)[12], the formed phase diagram shows a simple eutectic at 577 °C. Meanwhile, there is no report of metastable intermetallic compounds or glassy alloys in this binary. It has been known that above its solubility limit, Si would precipitate in particles or grains formed in the Al matrix[13]. On the other hand, Ge shows a higher solubility limit in Al with about 2% at a lower eutectic temperature of 420 °C. The diffusion coefficient of Ge in Al is higher than that of Si in Al at the investigated temperatures[14]. In a nanowire system, the understanding of the diffusion behavior and kinetics of this ternary are limited. As the ratio of surface area to the volume fraction increases, the surface energy could become a dominant factor. Consequently, the diffusion process would behave differently from bulk and thin film structures. In fact, due to the low solubility limits of Si in Al and vise versa, there is no report of Al thermal diffusion in Si NWs, while Al can easily diffuse in Ge NWs at low temperature (< 300 °C). This thermal diffusion process leads to the formation of a monocrystalline c-Al phase with an abrupt interface to the original Ge part and no intermediate phase is formed. The real time observation and diffusion kinetics of Al/Ge thermal exchange have been described in detail in the work of Luong et al.[15,16] and El Hajraoui et al.[17], where they propose that the rate of the thermal induced exchange reaction between the initial Ge NW and the Al is limited by Al volume diffusion, while Ge travels back into the Al metal contact by a surface diffusion mechanism. This interesting thermal diffusion process can be explained by the difference in diffusion coefficients, and we speculate that the driving force of the process is the fact that Ge can lower its energy by diffusing on a surface or grain boundary in Al metal when the system is heated. Additionally it was shown by Sistani et al.[18] that the selective propagation of Al in Ge-Si core shell NWs created a radial heterostructure with a thin Si shell from the original NW wrapped around the Al/Ge/Al axial heterostructure.

In this paper, we report in situ real time transmission electron microscopy (TEM) observations of a thermally assisted solid-state exchange reaction between an Al metal pad and a $Si_xGe_{1-x}$ alloy nanowire where we observe a reversible diffusion process, allowing to fabricate axial heterostructures starting from an initial homogeneous alloy NW. On heating, a reaction interface progresses in the NW where the Al metal replaces the original $Si_xGe_{1-x}$ NW. On controlled cooling, a Si rich region is progressively extending backwards inside the just converted Al NW section, going from the interface with the original NW toward the Al contact pad. The crystallographic and compositional analyses on the created NW heterostructures are carried out using geometrical phase analysis (GPA)[19,20] and energy dispersive X-ray spectroscopy (EDX) technique. 3D quantitative chemical reconstructions[21] of the heterostructure are presented to determine the distribution of chemical elements after the reaction process. We also present the result of the thermal reaction in passivated $Si_xGe_{1-x}$ NWs with a 20 nm $Al_2O_3$ shell, demonstrating a significant improvement of the interface quality. From the ex-situ and in-situ observations, we propose a hypothesis to interpret the possible kinetics for the formation of the $Si/Si_xGe_{1-x}$ NW heterostructure. A large electrical resistance difference is observed with or without the Si rich region, comparable to phase change materials (PCM) in memory devices[22,23]. Additionally, the created $Si/Si_xGe_{1-x}/Si$ NW heterostructure is also promising for near-infrared sensing applications[24–26].

### In-Situ TEM Observation of Al/Si$_{0.67}$Ge$_{0.33}$ NW Thermal Exchange Reaction

When two materials are brought in contact, depending on the thermal equilibrium of the system, inter-diffusion can take place at the contact interface due to the thermal vibration of atoms. Normally, ex-situ heating via rapid thermal annealing technique (RTA) is used to initiate and accelerate the exchange reaction benefiting from its simplicity. However, using such an ex-situ approach, no information is obtained on the mechanism and kinetics of the exchange reaction. In-situ heating experiments are therefore of high interest for a real time observation of the exchange reaction. In this study, as-grown NWs were dispersed on TEM temperature calibrated heater chips, aiming for a direct observation of the Al metal protruding in Si$_{0.67}$Ge$_{0.33}$ NWs. Fig. 1a shows the high-angle annular dark-field (HAADF) scanning TEM (STEM) image of an Al contacted Si$_{0.67}$Ge$_{0.33}$ NW lying over a hole before the annealing process. The NW diameter is about 150 nm with a length of 20 µm. The NW was contacted by a 200 nm thick Al rectangular pad and the heating experiment was conducted inside the TEM microscope. To better describe the diffusion behavior, we will discuss the in-situ experiment in two separate processes, i.e. (i) during the increase of the heating temperature when the temperature was slowly raised from room temperature to 580 °C (just above the eutectic temperature of Al/Si (577 °C)) with steps of 10 °C and (ii) during the controlled cooling down to room temperature where the temperature was kept constant at 560, 550 and 530 °C for observation.

*i) During the heating process*

Fig. 1a,c shows the contacted Si$_{0.67}$Ge$_{0.33}$ NW at the start of the experiment. When the temperature was increased following the temperature profile of Fig. 1e, the HAADF-STEM contrast indicated initiation of the thermal exchange reaction starting at 350 °C where a darker contrast associated to the Al metal starts to enter the Si$_{0.67}$Ge$_{0.33}$ NW through the NW surface underneath the Al contact pad. During the progression of the reaction interface, series of HAADF STEM images were taken with 0.787 s per frame to follow the diffusion behavior, presented in the Supporting Information **M1**. Fig. 1b shows the HAADF STEM image of the propagated NW when the heating temperature reached 580 °C after 45 min. The HAADF intensity is related both to the sample thickness and the mean atomic number of the elements. Since Si$_{0.67}$Ge$_{0.33}$ alloy density is heavier than that of Al, the brighter contrast corresponds to the initial Si$_{0.67}$Ge$_{0.33}$ NW, and the darker contrast to the entering Al metal. While the NW diameter is reduced slightly in the reacted region (from 150 nm to 146 nm), this reduction alone could not account for the strong HAADF contrast (see inset Fig. 1b). As indicated in Fig. 1b, the 'Reacted part' is the intrusion length of Al into the alloy NW of about 6.4 µm from the edge of the left Al contact pad. The insert in Fig. 1b shows a zoom of the reaction interface, which surprisingly demonstrates the presence of an intermediate contrast region between the original Si$_{0.67}$Ge$_{0.33}$ NW on the right and the reacted region on the left, referred to as the double interface region. For convenience, the left interface between the Al reacted part and double interface region is called the first interface and the one on the right between the double interface region and unreacted Si$_{0.67}$Ge$_{0.33}$ NW is called the second interface. The schematic illustrations of the contacted NW before and after annealing for 45 min are presented in Fig. 1c,d. The white arrow shows the direction of Al thermal diffusion during the heating process, and the orange segment demonstrates the formation of the double interface region. Presented in SI **M1**, the two interfaces show a relatively violent kinetic behavior along the NW growth direction during the increase of the temperature. The plot in Fig. 1f demonstrates the relative positions of the two interfaces during the heating. It can be observed that the interfaces do not advance smoothly, but rather advance in sub- 50 nm sized steps, that can be as large as tens of nanometers. Due to the random behavior of the interface jumps, the time axis is not linear but adapted to show the forward and backward movement of the first interface. A careful analysis of the in-situ experiment SI **M1** reveals that the two interfaces are halted at the same position along the NW and that the two interfaces can momentarily catch up, to be a single interface. Suddenly, the first interface moves backward to generate a new double interface segment along the NW axial direction, while the second interface can only halt or advance along the NW axis. The exchange reaction proceeds in this way and halts after a certain time when the temperature is constant. Then, the exchange reaction can be continued by increasing the temperature.

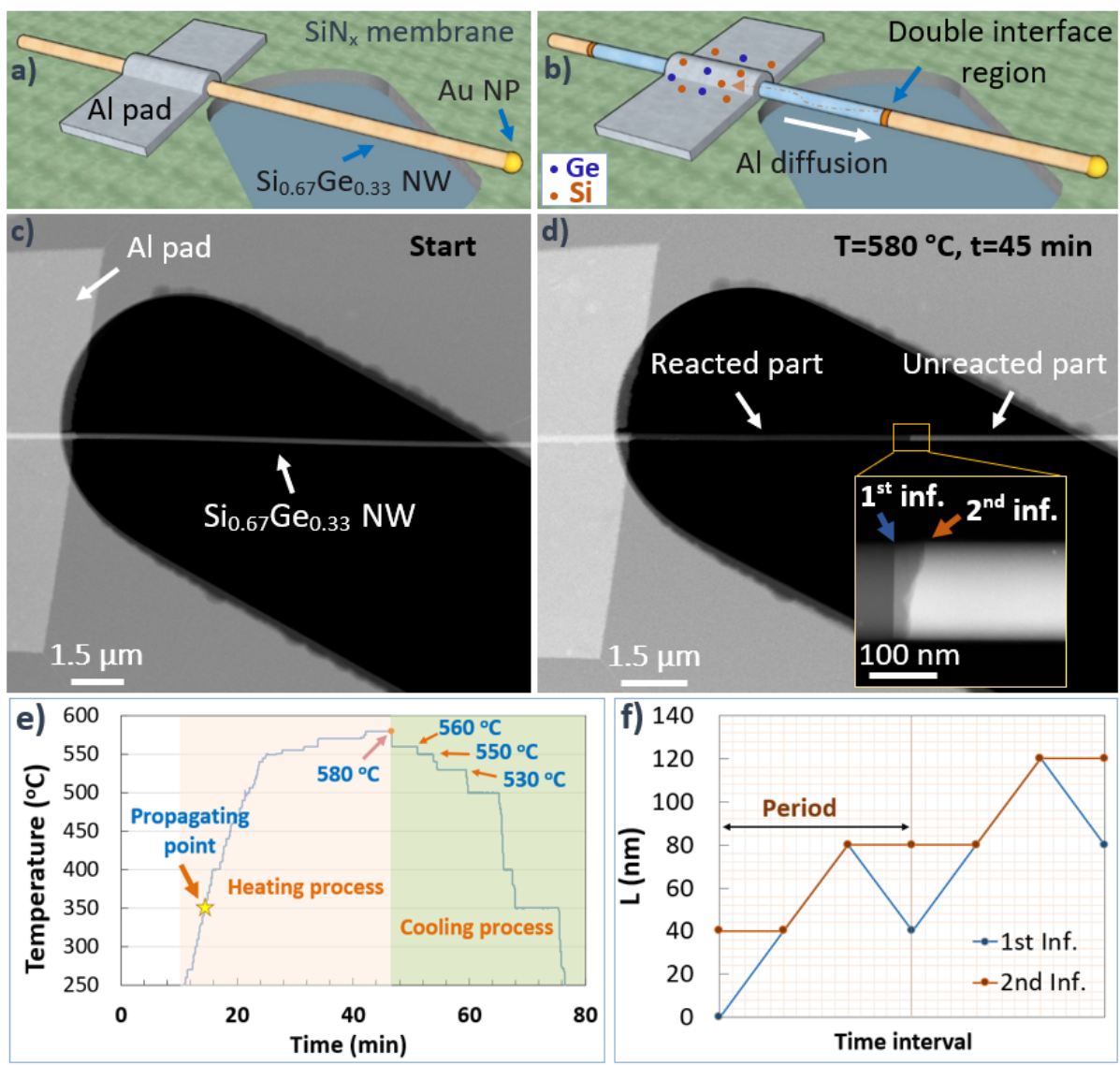

**Figure 1 | Real time observation of the thermal exchange between Al pads and a $Si_{0.67}Ge_{0.33}$ NW during the heating process. a-b)** Schematic illustrations of the contacted NW before and after annealing for 45 min. Si and Ge atoms are displayed in orange and blue colors, respectively. The white arrow in Fig. 1b shows the propagating direction of the reaction interface while the orange arrow shows the diffusion of Si and Ge atoms in the opposite direction. **c-d)** HAADF STEM images of the Al contacted $Si_{0.67}Ge_{0.33}$ NW crossing over the 6 μm x 23 μm hole of a $SiN_x$ membrane before and after the in-situ heating experiment. The diameter of the contacted NW is about 150 nm with a length of 20 μm. The insert in Fig. 1b shows the presence of a double interface region, sandwiched between the Al reaction part and unreacted $Si_{0.67}Ge_{0.33}$ part. **e)** The plot figure shows the heating temperature as a function of time during the heating and cooling process. **f)** The schematic illustration for the relative positions of the first and second interface during the propagation time. A real time propagation of Al/ $Si_{0.67}Ge_{0.33}$ interface is presented in the Supporting Information SI **M1**.

*ii) During the cooling process*

When the reaction interface had extended from the left to the right over a distance of 6.4 μm (at 580 °C), the heating temperature was slowly reduced down to room temperature. At some points (i.e., 560, 550 and 530 °C), the temperature was kept constant to investigate the evolution of the reaction interfaces (see Fig. 1e). Interestingly, the first interface was observed to extend in the reverse direction toward the left Al contact pad. Fig. 2a shows the HAADF-STEM image of the propagated NW at 560 °C during the cooling process. The insert shows a HR-STEM image with the corresponding Fourier transform (FT) on the [011] zone axis (taken at the yellow box), which indicates the presence of twin structure defects starting from the NW surface and running across the NW diameter along the $[1\bar{1}1]$ direction.

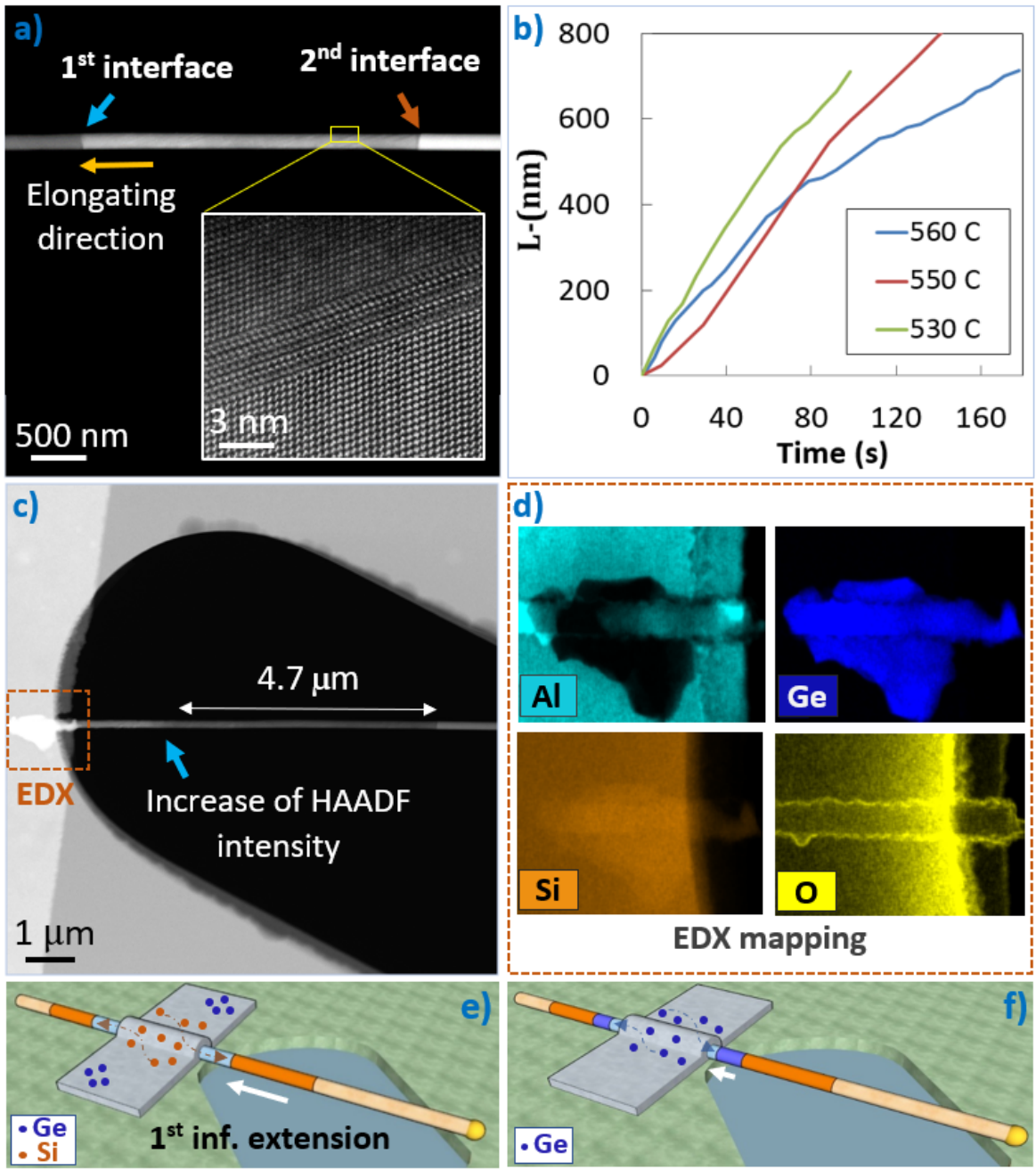

**Figure 2 | Real time observation of the thermal exchange between Al pads and a Si$_{0.67}$Ge$_{0.33}$ NW during the cooling process.**
**a)** HAADF-STEM image of a NW propagated at 560 °C during the cooling process, showing the movement of the first interface in the reverse direction toward the Al contact pad. The inset figure shows the zoom on the rectangular yellow box, demonstrating the formation of crystal defects in the newly formed NW region. **b)** The plot of the first interface diffusion length ($L$) in reverse direction as a function of time at three different cooling temperatures, showing a linear diffusion behavior. **c)** HAADF STEM image after 4.7 µm of backward propagation where an increase of HAADF intensity is observed in the created region, as well as the formation of a large crystal when the first interface reached the Al contact pad. **d)** EDX mapping on the created large crystal (indicated by the orange dash box) at the Al contact pad showing the main contribution of Ge in the crystal composition. The real time observation of the first interface during the cooling process (at 560 and 500 °C) are presented in the supporting information SI **M2** and SI **M3**, respectively. **e-f)** Schematic illustrations of the propagated NW during the cooling process, showing the backward diffusion of both Si and Ge atoms.

Fig. 2b shows the linear position dependence on time of the first interface diffusion length (*L*) in the reverse direction at three investigated temperatures (see legend Fig. 2b). It can be observed in Fig. 2b that the reverse reaction rate accelerates at lower temperatures as the steepest slope is observed for the lowest temperature of 530 °C, indicating that the reverse reaction is driven by the reduction of the temperature. The real time observation of the reverse reaction of the first interface (at 560 °C) is presented in the Supporting Information SI **M2**. The second interface was carefully investigated showing no modification in shape and position. When the first interface had extended over a distance of 4.7 µm back toward the left Al contact pad, there was an enhancement of the HAADF intensity in the reverse reacted region until the reaction interface reached the Al contact pad, where a large crystal was then formed on the Al reservoir (see Fig. 2c). The real time observation of the backward diffusion with the increase of the HAADF intensity is presented in the supporting information SI **M3** (at 500 °C). EDX mapping acquired on the created crystal (Fig. 2d) shows the main contribution of Ge in the crystal composition. The Si- signal appears very weak and could be an artifact due to the scattered X-rays from the $Si_3N_4$ membrane. Fig. 2e-f show the schematic illustrations for the backward diffusion of first Si and later Ge atoms during the cooling process. Si atoms, being solute in the Al contact pad after the heating procedure, return to the NW during the cooling and fill the NW volume pushing the 1$^{st}$ interface in the reverse direction (Fig. 2e) which is then followed by Ge atoms (Fig. 2f).

## Structural and Compositional Analysis on Created Heterostructures

To understand the mechanism of the exchange reaction, it is necessary to characterize the distribution of elements within the created structure. For these analyses, geometrical phase analysis (GPA) and quantitative EDX with a 3D reconstruction on different reacted parts were performed. For the experiments, as-grown $Si_{0.67}Ge_{0.33}$ NWs were contacted on the 200 nm thick $Si_3N_4$ membrane and annealed ex-situ via RTA at 400 °C for 20 s and then rapidly cooled down to room temperature during 4 min. The HAADF STEM image of a contacted NW crossing over a 2 µm x 8 µm hole after the thermal treatment is presented in the supporting information SI **S1**a. Analysis of the lattice spacing using GPA indicated that the formed segment with intermediate contrast between the created Al part and original $Si_{0.67}Ge_{0.33}$ NW contains mostly Si.

To obtain more precise chemical characterization, Fig. 3 shows EDX quantification on each region of the created heterostructure (indicated by the dash yellow box in Fig. S1b). The distribution of Al, Si, Ge and O elements are displayed in turquoise, orange, blue and yellow, respectively. Fig. 3a presents a line-scan crossing the double interface region from the Al reacted part to the unreacted $Si_{0.67}Ge_{0.33}$ part. The normalized concentration profile in atomic percent (at. %) was extracted using QUANTAX-800 software from BRUKER, showing a transition from an Al part to a Si-rich region and then $Si_{0.67}Ge_{0.33}$ original part. It is apparent that the double interface region is mainly made of Si atoms. In addition, we observe a small gradient of Ge concentration from the first interface to the second interface, which is coherent with the HAADF intensity variation found in Fig. S1b. The 3D reconstructions of the NW cross-section on each particular region (using the modeling method[21]) are presented in Fig. 3b-d. Fig. 3b presents the mapping, radial line-profile and 3D reconstruction of the original $Si_{0.67}Ge_{0.33}$ NW. Firstly the line-scan profile in the cross-section direction demonstrates the presence of a homogenous $Si_{0.67}Ge_{0.33}$ core, which is then covered by an oxide shell. Quantitatively, the as-grown NW has a $Si_{0.67}Ge_{0.33}$ core with the asymmetric diagonals of 74.3 nm and 65.4 nm. It is covered by 0.5 nm of thin Ge-rich shell (~43%) and 1 nm $SiO_2$ shell. EDX quantification for the Si-rich region is shown in Fig. 3c. In the core part, there is an enrichment of the Si concentration with respect to the original part (67% Si) up to ≥ 90%. This region is also composed of small proportions of 5 - 6% Ge and 0 - 2% Al. It should be noted that the finding of ~2%Al in the core can be an artifact due to scattered X-rays of Al from the large Al contact pads. However, Al atoms might also diffuse into structure defects present in the Si rich segment (as shown in Fig. 2a). The outer shells are composed of a thin shell (about 1 nm) containing some Ge (8%) and then a mixture of $Al_2O_3$ and $SiO_2$ shell of about 2 nm. The presence of Al atoms in the shell parts can be attributed to the hypothesis that the Al reaction interface had reached the $Si_{0.67}Ge_{0.33}$ interface and then moved backwards creating the Si-rich region (see also SI **M1**, **M2** and **M3**). Fig. 3d shows the compositional analysis on the reacted part of the NW. The converted region has an Al core containing a noticeable 4%Si and small percent of Ge (about the quantification limit). In the literature[12], the solubility limit of Si in Al is below 2%. Since the investigated region is near the $Si_3N_4$ membrane, a small contribution of Si X-rays scattered from the $Si_3N_4$ membrane is un-avoidable, which will contribute to this quantification result. The outer shells consist of about 1 nm Ge containing shell and 1-2 nm of mixed $Al_2O_3$ and $SiO_2$ shell. It can be observed that the dimensions of especially the Si rich and also the Al converted region are smaller than the original NW dimensions.

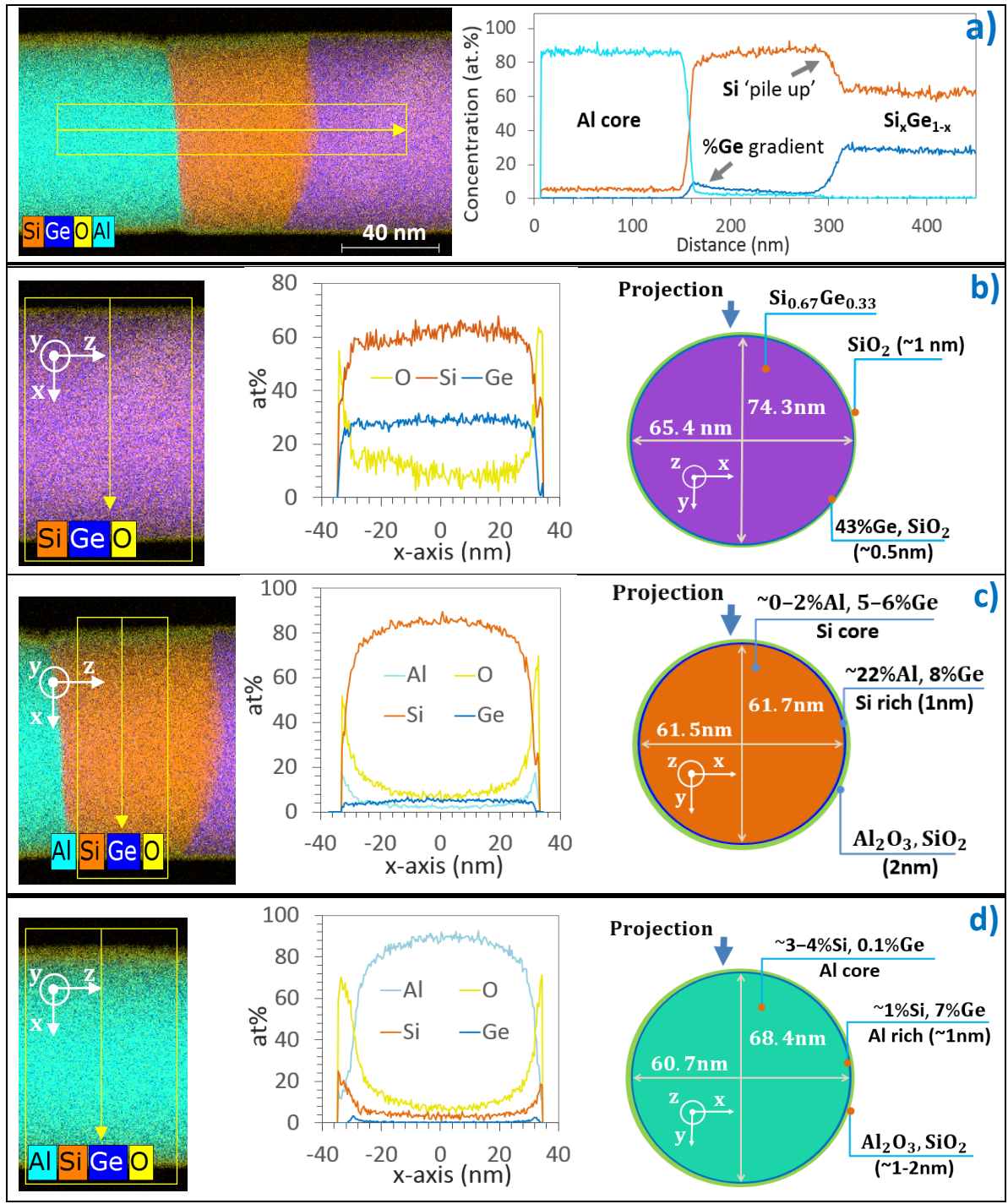

**Figure 3 | Compositional analysis on the created heterostructure using quantitative energy dispersive X-ray spectroscopy technique.** EDX mapping on the created heterostructures shown in Fig. S1b (the dash yellow box). **a**) EDX hyper-map and line-scan profile crossing the heterostructures from the reacted, double interface region and unreacted $Si_{0.67}Ge_{0.33}$ part. **b-d**) EDX quantification on three different parts of the heterostructures, showing the chemical map, chemical profile and 3D reconstruction, respectively. The vertical arrow indicates the projection direction of the elliptic reconstruction model.

### Influence of the Cooling Speed on the Si-rich Segment Length

From the real time observation of the Si-backward diffusion presented in Fig. 2, it appeared that the cooling rate was the main parameter determining the extension length of the Si-rich segment. To confirm this prediction, we separated a similar set of fabricated Al contacted $Si_{0.67}Ge_{0.33}$ NWs into two parts and applied a thermal treatment (RTA) with two different recipes with a fast and slow cooling step respectively. As can be seen in the SI **S2**, indeed the Si-rich segment length is much longer with a slow cooling step. This experiment interestingly demonstrates the possibility to tailor the $Si/Si_xGe_{1-x}/Si$ heterostructure size by controlling the temperature ramp during the heating and cooling process.

### Influence $Al_2O_3$ Passivation Shell on the Interface Shape

We have seen a clear influence of the NW diameter on the $Si_{0.67}Ge_{0.33}$ interface shape, see SI **S3**. Particularly, small NWs with diameters ranging from 60 to 100 nm show the formation of a clean Si-rich/$Si_{0.67}Ge_{0.33}$ interface with a straight or more often observed, convex shape (Fig. S3a). Larger NW diameters (150 – 250 nm) however show the presence of a very rough interface with different facets (shown in Fig. S3b).

From the in-situ observation presented in Fig. 1b, where it was observed that both interfaces move in relatively large steps and both interfaces halt at the same positions along the NW, we speculated that the interfaces can be trapped at specific surface locations and therefore the formation of the interface shape will be strongly influenced by the surface quality. To verify this hypothesis, we cleaned the NW surface by hydriodic acid (HI) and immediately passivated the as-grown NWs with a 20 nm $Al_2O_3$ shell [using atomic layer deposition (ALD) at 250 °C]. The NWs were then contacted on both ends with 200 nm thick Al pads and deposited on $Si_3N_4$ membranes and TEM calibrated heater chips for ex-situ and in-situ annealing experiments, respectively. For the ex-situ heating experiment, the contacted NWs were annealed at 450 °C for 10 s and rapidly cooled down to room temperature. Fig. 4a shows the HAADF STEM image of a contacted 77 nm thick $Si_{0.67}Ge_{0.33}$ NW with a 20 nm $Al_2O_3$ passivation shell after the thermal treatment, showing the Al conversion length of about 650 nm and Si-rich segment of 14 nm. Fig. 4b shows the magnified image taken in the blue box (Fig. 4a), showing the alignment of $Si_{0.67}Ge_{0.33}$ (111) // Si rich (111) // Al (111) planes. Again, we can observe a brighter HAADF contrast between the Si rich region and Al reacted part (about 5 atomic planes), which is attributed to a locally increased Ge content.

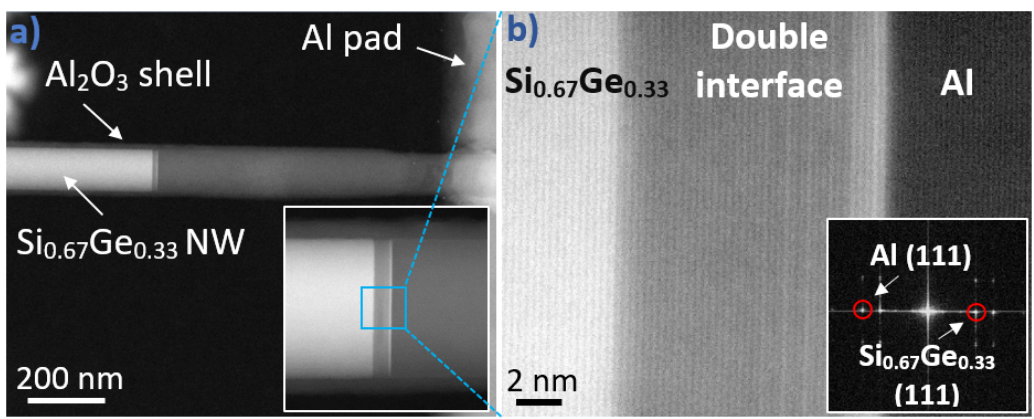

**Figure 4 | Improvement of interface shape by using passivated NWs. a)** HAADF STEM image of propagated NW showing the formation of Al/Si rich/$Si_xGe_{1-x}$ heterostructure. **b)** HR HAADF STEM image with the corresponding FT image taken on the blue box, showing an epitaxial alignment of Al/Si rich and Si rich/$Si_{0.67}Ge_{0.33}$ interfaces on the [111] reflection. Also see the real time observation of the Al thermal diffusion in passivated NWs shown in supporting information SI **M4** and Fig. **S4**

We have also performed the in-situ heating experiment in two coalesced and passivated $Si_{0.67}Ge_{0.33}$ NWs (see SI **S5** and **M4**). As can be seen from the movie, the diffusion process was smoother and the $Si_{0.67}Ge_{0.33}$ interface appeared very sharp. Compared to the unpassivated NW (Fig. 2a), the crystalline quality of the Si-rich region is improved, as no twin defects were observed in the passivated NWs. These results confirm the critical role of NW surface quality on the diffusion behavior and kinetics of the exchange reaction. Probably, the NW surface quality had been significantly improved after the NW surface cleaning by wet etching process and the protection of 20 nm $Al_2O_3$ passivation shell.

### Transport properties through the created heterostructures

For device applications, it is of crucial importance to investigate the electrical transport property of the created heterostructures. We have therefore performed electrical measurements on the $Si_{0.67}Ge_{0.33}$ NWs before and after the thermal diffusion. Fig. S6a shows the BF STEM image of a contacted NW after the thermal reaction via RTA. The exchange reaction has taken place from two sides of the contacted NW (190 nm in diameter), leaving a remaining unreacted $Si_{0.67}Ge_{0.33}$ segment of about 1.75 µm. The plot in Fig. S6b shows the IV characteristics of the contacted NW before and after the thermal propagation. Firstly, we can see that the current passing through the un-doped $Si_{0.67}Ge_{0.33}$ NW is very low, in the order of $10^{-10}$ A for a biasing voltage of 1 V. After the metal intrusion, the contact resistance had increased significantly so that the measured current dropped two orders of magnitude. Potentially with the presence of the Si-rich region between the reacted and unreacted part, which has a large bandgap energy, the flowing current is blocked by the band offset at the $Si/Si_{0.67}Ge_{0.33}$ interface, causing the drop of the measured current. The possibility to modulate the resistance is promising for applications as phase change materials in an all-crystalline system, avoiding known challenges as chemical segregation[27–29].

### Repetitious cycle property

The diffusion behavior of this ternary system combined with the change in resistivity may be promising for phase change memory applications[22,23]. Therefore, we have tested the cyclability on an Al contacted un-passivated $Si_{0.67}Ge_{0.33}$ NW (having a diameter of 210 nm) by exposing it to alternating heating and cooling cycles around the eutectic temperature of Al/Si, as shown in SI **M5**. Interestingly, it is possible to remove and re-produce the Si rich segment for several cycles while maintaining the NW geometry and unreacted $Si_{0.67}Ge_{0.33}$ part. However, the Si rich segment length appeared shorter after each cycle, which is attributed to loss of Si atoms which had moved out and hadn't returned to the NW. It is therefore necessary to optimize the temperature window during the cooling process to stabilize the number of returning Si atoms for a precise control of the Si rich segment length.

### Discussion

Based on above observations of the thermal diffusion behavior, we propose some hypotheses to interpret the kinetics of the thermal reaction forming $Al/Si/Si_{0.67}Ge_{0.33}$ heterostructures. Recalling the diffusion behavior shown in the SI **M1**, the two interfaces moved forward catching up with each other after a certain time, and then the $Si_{0.67}Ge_{0.33}$ interface suddenly stopped while the Si rich interface moved backward to generate a new Si rich region. Firstly, the step-wise propagation of the $Si_{0.67}Ge_{0.33}$ interface may be explained due to the presence of defects or roughness on the NW surface. Since as-grown NWs were exposed to air, the NW surface was strongly oxidized resulting in the formation of surface defects. The EDX quantification has shown the presence of an about 1 – 2 nm thick $SiO_2$ shell around the NW. Therefore, the trapping and detrapping of the reaction interface at these defects may cause the discontinuous propagation of the $Si_{0.67}Ge_{0.33}$ interface and the presence of several facets. This hypothesis was corroborated by the formation of a clean and sharp $Si_{0.67}Ge_{0.33}$ interface when passivating the NWs by a 20 nm $Al_2O_3$ shell. Then, the Si rich interface moved in the reverse direction after the $Si_{0.67}Ge_{0.33}$ interface had stopped. From Fig. 3d, we observed about 4% of Si content in the Al reacted part (higher than the solubility limit of Si in Al, 1.64%[12]), this part may be saturated by Si atoms. The saturation of Si content in the diffusion channel results in the precipitation of Si at the $Si_{0.67}Ge_{0.33}$ interface, forming the Si rich region in an in-equilibrium state. The Si rich region then acts as a barrier layer preventing Ge atoms of the original $Si_{0.67}Ge_{0.33}$ part from moving toward the Al reservoir, and hence the $Si_{0.67}Ge_{0.33}$ interface is blocked during the presence of the Si rich region. After a certain time, when the Si content in the diffusion channel (Al core and surface in the reacted part) drops below the solubility limit, the Si rich segment disappears and the $Si_{0.67}Ge_{0.33}$ interface again starts a new cycle of propagation. It is also observed that there is a gradient of Ge content in the Si rich region with a maximum of Ge at the Al/Si rich interface. This may indicate that forming an interface between monocrystalline Al and Si is energetically not allowed at the experimentally accessible temperature. Therefore, to decrease the energy of the interface, the Ge content must increase at the Al interface. This would also explain why the exchange reaction was never observed in the Al/Si binary couple.

From the SI **M2**, when the temperature was slowly reduced, the Si rich interface was observed to move in the reverse direction toward the Al contact pad, consequently extending the Si rich region. Since the interface with the original $Si_{0.67}Ge_{0.33}$ NW didn't show any change during extension of the Si rich region, it is expected that Si atoms that previously moved out of the NW and into the grain boundaries or surface of the Al reservoir are now defusing back and reconstitute the NW cross-section. From literature [30–32], this phenomenon can be explained by the solidification process of the Al/Si binary system (with eutectic temperature of 577 °C). The lowering of the temperature below the eutectic temperature leads to a considerable drop of the Si solubility in the Al reservoir, resulting in the precipitation of Si atoms. The unreacted $Si_{0.67}Ge_{0.33}$ part can be a reasonable nucleation point for the precipitation due to the high Si content. During the elongation of the Si rich region, we have observed the shrinkage of the NW diameter that may occur due to the replacement of Ge atoms by Si atoms with smaller radius, making the crystal structure become more compact. Reduction of the NW diameter could also partly be explained by the fact that the NW volume does not necessarily have to be filled with Al or Si (and Ge) atoms exactly to the level it was prior to any reaction. After the backward diffusion of the Si atoms, we observed the increase of HAADF intensity evidencing returning Ge atoms, also demonstrated by the formation of a crystal on the Al pad when the retreating reaction interface reached the Al pad. From the EDX mapping of the crystal presented in Fig. 2d we find that the crystal is mainly composed of Ge atoms. The Si signal at the Ge crystal appears very weak, which indicates an exhaustion of Si content in the Al contact reservoir. The backward diffusion of Ge atoms after the exhaustion of Si atoms is explained by the lower Al/Ge eutectic temperature (420 °C) compared to Al/Si (577 °C). The backward diffusion of Si followed by Ge atoms is a remarkably interesting phenomenon that has not been reported in literature, and has not been observed in the binary Al/Ge NW system[17]. These results demonstrate that the NW geometry guides the recrystallization of a supersaturated element (in this case Si and Ge) in a controllable and repeatable fashion, which may also apply to other material couples.

Unlike the diffusion behavior in large NW diameters (≥ 150 nm) resulting in the formation of a rough $Si_{0.67}Ge_{0.33}$ interface, the created $Si_{0.67}Ge_{0.33}$ interfaces in small NW diameters (≤ 100 nm) are typically clean with a convex shape. Fig. 5a shows the HAADF STEM image of a propagated $Si_{0.67}Ge_{0.33}$ NW lying over a hole on the $Si_3N_4$ membrane (**a**) and a zoomed image on the top part of the reaction interface (**b**). Magnifying the reaction interface shown in Fig. 5b, we can see a straight interface at the center and a bending shape when it comes to the interface edge. From this interface shape it is expected that the exchange reaction does not start from the surface, but nucleates in the NW core and extends to the surface. This can be the reason why the created interfaces are mostly abrupt and clean without any facet. Fig. 5c-d presents the schematic illustrations of Al, Si and Ge diffusion direction during the heating (c) and cooling (d) process. During the forward diffusion of the $Si_{0.67}Ge_{0.33}$ interface, Al atoms from the contact pad will move through the converted Al region to exchange with the Si and Ge atoms at the interface (**path 1**), starting from the center and spreading out toward the interface edge. From the results of the EDX quantification presented in Fig. 3, in the unreacted $Si_{0.67}Ge_{0.33}$ part, Si and Ge surface atoms make bonds with oxygen atoms forming the stable $SiO_2$ shell (with a small fraction of $GeO_2$ shell since the Si/Ge ratio is ~0.67/0.33). Whereas the freshly created Al part is oxidized creating an $Al_2O_3$ shell. Considering the different bond dissociation energy (enthalpy) for Si-O (799.6 ± 13.4 kJ/mol) and Al-O (501.9 ± 10.6 kJ/mol) [33], the $SiO_2/Si_{0.67}Ge_{0.33}$ interface is more stable than the $Al_2O_3/Al$ interface. Therefore, when approaching the NW surface, the exchange reaction is decelerated at the interface edge causing the formation of a convex interface. A similar argument was raised in the paper of Chou et al.[34], where they interpreted the formation of a convex interface between Si and Co. Turning back to Fig. 5c, after being replaced by Al atoms, Si and Ge atoms from the reaction interface pass through a surface diffusion channel toward the Al contact pad (**path 2**) where they diffuse on surfaces and grain boundaries of the large Al reservoir. Previously, kinetic experiments in thin Ge NWs with Al contacts indicated that the reaction rate is limited by self-diffusion of Al (volume diffusion through the created segment), while Ge can diffuse back to the reservoir by surface diffusion (El Hajraoui et al.[17]). When the temperature is reduced (Fig. 5d), due to the solidification process, Si atoms from the Al reservoir now return and precipitate at the $Si_{0.67}Ge_{0.33}$ interface, pushing the Si rich interface in the reverse direction (**path 1**). Al atoms at the reaction interface are replaced by Si and Ge atoms, forcing the Al atoms back to the Al contact pad (**path 2**).

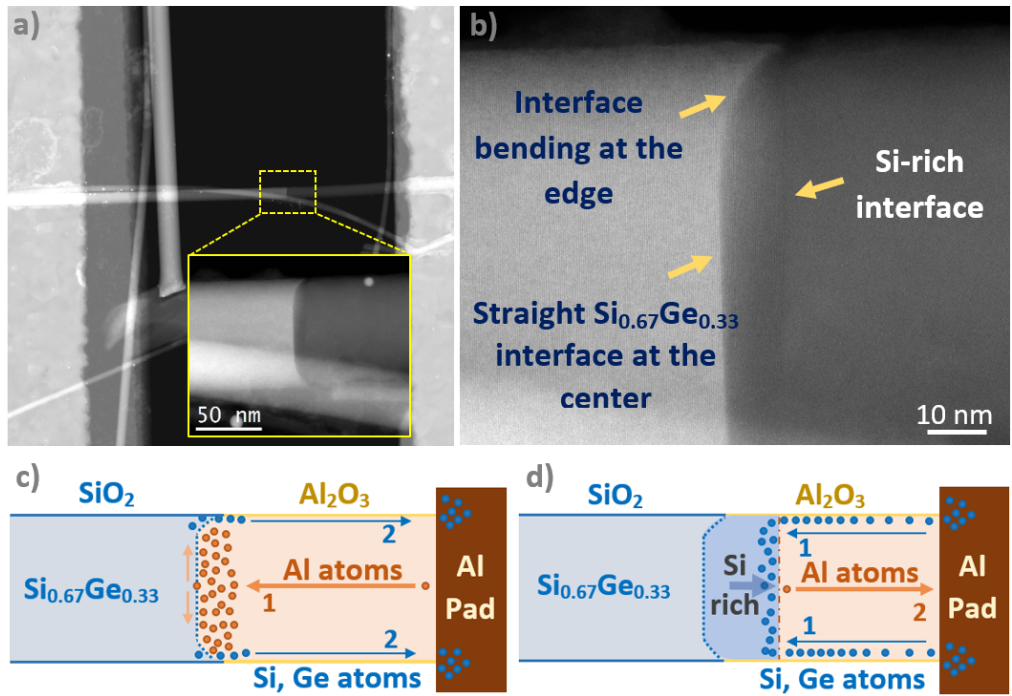

**Figure 5 | Diffusion mechanism in small NW diameter. a**) HAADF-STEM image of Al contacted $Si_xGe_{1-x}$ NW (83 nm in diameter) after being annealed at 400 °C for 20 s, 400 to 300 °C for 30 s and cooled down to room temperature during 4 min. The inset figure shows the zoom on the reaction interface having a convex shape. (**b**) The magnified image on the top part of the reaction interface showing a straight interface at the center and a bending shape at the NW edge. **c-d**) Schematic illustrations of Al, Si and Ge diffusion direction during the forward propagation of the $Si_xGe_{1-x}$ interface and backward diffusion of the Si-rich interface, respectively.

In the case of passivated NWs, due to the presence of the pre-deposited $Al_2O_3$ shell on both sides of the reaction interface, the interfacial energy difference between the unreacted segment ($Al_2O_3/Si_{0.67}Ge_{0.33}$) and reacted segment ($Al_2O_3/Al$) has been reduced. Therefore, the reaction rate is unaffected by the interfacial energy difference on both sides of the reaction interface, resulting in the formation of a flat $Si_{0.67}Ge_{0.33}$ interface (see Fig. 4).

## CONCLUSION

In summary, the Al- (Si, Ge) NW thermal exchange reaction was investigated in detail via ex-situ and in situ heating techniques. The incorporation of Al metal in $Si_{0.67}Ge_{0.33}$ alloy NWs results in the formation of a Si-rich region (≥90%), sandwiched between the reacted and unreacted part of the NW. When reducing the heating temperature, we have observed a linear extension of the Si rich segment length in the reverse direction back to the Al contact pad. In small NW diameters, the interfacial energy between the core and oxide shell on both sides of the reaction interface becomes more significant and governs the exchange reaction. The clean-convex $Si_{0.67}Ge_{0.33}$ interface results from the deceleration of the diffusion rate from the center to the edge of the NW due to the lower interfacial energy of the $SiO_2/Si_{0.67}Ge_{0.33}$ interface compared to that of the $Al_2O_3/Al$ interface. With a pre-deposited $Al_2O_3$ shell, the interfacial energies on both sides of the reaction interface are reduced significantly, which results in the formation of flat atomically abrupt $Si_{0.67}Ge_{0.33}$ and Si rich interfaces and much smoother advancement of the reaction interface. Then during the cooling process, the linear extension of the Si and Ge atoms are explained by the solidification process of Al- (Si, Ge) ternary system. Slowly decreasing the temperature induces the precipitation of first Si and then Ge atoms, starting at the unreacted $Si_{0.67}Ge_{0.33}$ part. Current-voltage characteristics without (on-state) and with the presence of the $Si/Si_xGe_{1-x}/Si$ heterostructure (off-state), show a current drop of two orders of magnitudes ($I_{on}/I_{off} = 10^2$), respectively. This ratio is expected to be improved when employing doped $Si_xGe_{1-x}$ alloy nanowires to raise the on-state current. In summary, these findings show the possibility to produce tunable $Al/Si/Si_xGe_{1-x}$ axial heterostructures with integrated contacts, created in a single fabrication step. These structures may show a great potential for phase change memories as well as near infrared

(NIR) sensing applications since $Si_xGe_{1-x}$ alloys show an interesting flexible bandgap that can be tuned in the 1.3 µm to 1.55 µm telecommunication window, all in a system compatible with current silicon-based technology. Moreover, the observed reversible thermal diffusion of the metal, certainly due to the NW system confining semiconductor and metal in an oxidized shell, may be a concept that can be extended to other material couples.

## Methods

In this work, we have conducted the thermal diffusion of Al metal into three different $Si_xGe_{1-x}$ alloy stoichiometries (i.e. x= 0.05, 0.12 and 0.67). However, for Al diffusion in low composition Si NWs (x= 0.05 and 0.12) no clear difference with the pure Ge NW system was observed, and a strikingly different behavior was observed in the NWs with x= 0.67. Therefore, we decided to focus on the $Si_xGe_{1-x}$ NWs (x= 0.67) in our experiments. The growth of $Si_{0.67}Ge_{0.33}$ NWs was performed by chemical vapor deposition method (CVD) via the VLS growth mode, using silane and $GeH_4$ gasses as precursors and gold as catalyst on a Si(111) substrate. The NWs were growth along the [111] direction. The fabricated NWs have a strong variation in diameter, ranging from 60 to 250 nm. For the experiments, as-grown NWs were either being used directly to perform metal contacts or a surface passivation was applied by cleaning the NW surface by dipping in diluted hydriodic acid (HI) for 5 s and immediately passivating by a 20 nm $Al_2O_3$ shell (using atomic layer deposition (ALD) at 250 °C). Aiming for in-situ heating experiments, $Si_{0.67}Ge_{0.33}$ NWs were dispersed on calibrated heater chips from DENSsolution company[35], which contain several 6 µm x 23 µm holes on the $SiN_x$ membrane and have the possibility to raise the temperature up to 1300 °C within a few seconds. We then selected NWs lying over the holes and contacted both sides by a pair of Al rectangular pads. Prior to the deposition of Al metal layer, NWs with the 20 nm $Al_2O_3$ shell were immersed in buffered hydrofluoric acid - BOE 7:1 (HF : $NH_4F$ = 12.5 : 87.5%) for 40 s to remove the $Al_2O_3$ shell in the contact regions and then dipped in diluted hydriodic acid (HI) for 5 s to etch the native $GeO_2$ shell. After that, the samples were cleaned by soft Ar plasma for 15 s and coated by a 200 nm Al thick layer using electron beam evaporation (with the purity of 99.995% and in vacuum at a pressure lower than $10^{-6}$ Torr). The samples were lifted off in Acetone solution overnight to obtain the final devices. For temperature calibrated in-situ heating experiments, the samples were heated inside the TEM microscope using a commercial DENSsolution six contact double tilt TEM heating holder. The in-situ heating process started from room temperature and the temperature was gradually increased with 10 °C steps until reaching 580 °C. We then slowly reduced the heating temperature and stopped at certain temperatures for observation. In this study, $Si_{0.67}Ge_{0.33}$ NWs were also contacted on home-made 200 nm $Si_3N_4$ membranes for ex-situ heating experiments using rapid thermal annealing (RTA). The fabrication of the $Si_3N_4$ membranes is reported elsewhere[36]. For ex-situ heating experiments, several contacted NWs can be propagated at the same condition giving more statistics on the reaction behavior. To conduct the thermal exchange reaction, specimens were annealed in a temperature range of 400 to 450 °C in $N_2$ atmosphere, and rapidly or slowly cooled down to room temperature. The RTA experiments were done in a Jipelec™ JetFirst RTP Furnace[37]. The data of crystallographic and compositional analysis were collected using a FEI Titan Themis microscope equipped with a probe Cs corrector and SuperX EDX (4 SDDs) detectors working at 200 kV. High angle annular dark field (HAADF) scanning TEM (STEM) was performed with a beam convergence angle of 20.7 mRad and electron beam current of 96 pA.

## ABBREVIATIONS:

CVD: chemical vapor deposition; VLS: vapor-liquid-solid; NWs: nanowires; ALD: atomic layer deposition; STEM: scanning transmission electron microscope; RTA: rapid thermal annealing; HAADF: high-angle annular dark-field imaging; EDX: energy dispersive X-ray spectroscopy; GPA: geometric phase analysis; NIR: near infrared.

## ASSOCIATED CONTENT

* The Supporting Information and Movies are available free of charge on the website:

https://drive.google.com/open?id=1U9u5DtOuoDXe-xoBYMCOt-IggH2s7-dB

SI M1 shows the real time propagation of Al metal into a 150 nm $Si_{0.67}Ge_{0.33}$ NW lying over a hole on the SiNx membrane (calibrated heater chip from DENSsolution company) during the heating process. SI M2 shows the backward diffusion of the first interface during the cooling process (at 560 °C). SI M3 shows the increase of the HAADF contrast and formation of a crystal when the first interface reaches the Al contact pad (at 500 °C). SI M4 shows the real time observation of the exchange reaction on two passivated $Si_{0.67}Ge_{0.33}$ NWs. SI M5 shows the real time propagation in an Al contacted un-passivated $Si_{0.67}Ge_{0.33}$ NW (having a diameter of 210 nm) during an alternating heating and cooling process around the eutectic temperature of Al/Si. Supporting information S1 shows the structure analysis of the created heterostructures using geometrical phase analysis. Supporting information S2 shows the comparison of the Si-rich segment length ($L$) in propagated NWs with respect to two different cooling processes. Supporting information S3 shows the influence of NW diameter on the $Si_{0.67}Ge_{0.33}$ interface shape. Supporting information S4 shows the influence of the passivation shell on the formation of the $Si_{0.67}Ge_{0.33}$ interface shape. Supporting information S5 shows the real-time observation of thermal exchange reaction in a double passivated $Si_{0.67}Ge_{0.33}$ NWs. Supporting information S6 shows the IV characteristics measured on the contacted NW before and after the thermal propagation.

* Competing interests:

The authors declare that they have no competing interests.

## AUTHOR INFORMATION

Corresponding Authors

*E-mail: martien.den-hertog@neel.cnrs.fr

## ACKNOWLEDGMENTS

We acknowledge support from the Laboratoire d'excellence LANEF in Grenoble (ANR-10-LABX-51-01). We benefitted from the access to the Nano characterization platform (PFNC) in CEA Minatec Grenoble and Nano-fab from institute NEEL, Grenoble. This project has received funding from the European Research Council (ERC) under the European Union's Horizon 2020 research and innovation programme (grant agreement N° 758385) for the e-See project. The authors gratefully acknowledge financial support by the Austrian Science Fund (FWF): project No.: P28175-N27.